\documentclass[twocolumn,showpacs,prl]{revtex4} 
\usepackage{graphicx}
\usepackage{natbib} 
 
\newcommand{\ds}{\displaystyle} 
 
\newcommand{\be}{\begin{equation}} 
\newcommand{\ee}{\end{equation}} 
\newcommand{\beq}{\begin{eqnarray}} 
\newcommand{\eeq}{\end{eqnarray}}

\newcommand{\w}{\omega} 
\newcommand{\W}{\Omega} 
\newcommand{\g}{\gamma} 
\newcommand{\G}{\Gamma}

\newcommand{\ket}{\rangle}

\newcommand{\bnn}{\begin{eqnarray*}} 
\newcommand{\enn}{\end{eqnarray*}} 
 
\newlength{\textwidthm} 
\begin{document}

\title{ 

Electromagnetically induced coherent backscattering
} 
 
\author{ 
    Yuri V. Rostovtsev$^{1}$, 
    Zoe-Elizabeth Sariyanni$^{1}$, 
and 
    Marlan O. Scully$^{1,2,3}$ 
} 
 
\affiliation { 
${^1}$Institute for Quantum Studies and Department of Physics, 
Texas A\&M University, 
TX 77843 \\ 
%
${^2}$Princeton Institute for the Science and Technology of Materials 
and Department of Mechanical \& Aerospace Engineering, 
Princeton University, NJ 08544 \\ 
${^3}$Max-Planck-Institute f\"ur Quantenoptik, D-85748 Garching, Germany 
} 
 
\date{\today} 
 
\begin{abstract} 
\vskip12 pt 
 
We demonstrate a strong coherent backward wave oscillation using 
forward propagating fields only. 
This is achieved by applying laser fields 
to an ultra-dispersive medium 
with proper chosen detunings  
to excite a molecular vibrational coherence that corresponds 
to a backward propagating wave. 
The physics then has much in common with propagation of ultra-slow light. 
Applications to coherent scattering and 
remote sensing are discussed. 
 
\end{abstract} 
 
\maketitle

Quantum coherence \cite{eit, book} 
has been shown to result in many counter-intuitive phenomena. The 
scattering via a gradient force in gases~\cite{harris-prl}, 
the forward Brillouin scattering in ultra-dispersive resonant 
media~\cite{matsko00prl,matsko01prl}, electromagnetically induced transparency
\cite{harris, merriam,xio,coussement02prl}, slow light
\cite{matsko,hemmer,hau,kash}, Doppler broadening elimination~\cite{ye}, 
light induced chirality in nonchiral medium~\cite{sau05prl}, 
a new class of entanglement amplifier~\cite{cel} based on correlated
spontaneous emission lasers~\cite{Scully,schleich}  
and the coherent Raman scattering enhancement via maximal 
coherence in atoms~\cite{JainHarris} and biomolecules~\cite{scully02pnas,ZY04jmo} are a few examples 
that demonstrate the importance of quantum coherence. 
 
In this Letter, we predict strong coherent backward scattering via 
excitation of quantum coherence between atomic or molecular levels. 
The developed approach can also be used to control the direction of 
the signal generated in coherent Raman scattering 
and other four-wave mixing (FWM) schemes.

\begin{figure}[tb] 
\center{ 
\includegraphics[width=7cm]{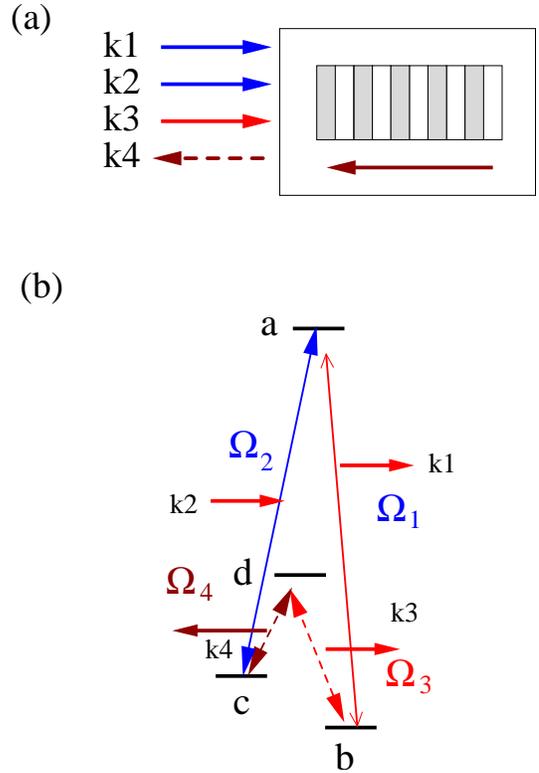} 
} 
\caption{\label{4wm} 
(a) Co-propagating fields 1 and 2 
induce coherent grading inside the 
medium. 
The field 3 propagating in the same direction 
will be scattered in the opposite direction because the coherence excited by 
fields 1 and 2 is propagating in the opposite direction (see Fig.\ref{kw}). 
Level scheme, double-$\Lambda$ (b), 
for implementation of coherent back scattering. 
} 
\end{figure}

\begin{figure}[tb] 
\center{ 
\includegraphics[width=7.8cm]{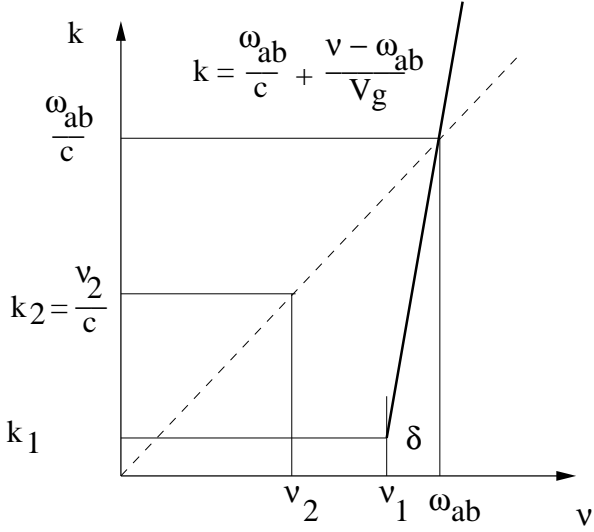} 
} 
\caption{\label{kw} 
Dispersion $k(\nu)$ of ultra-dispersive medium. 
Choosing $\delta = \nu_1 - \w_{ab}= -V_g\w_{cb}/c$, 
we can have $k_1 - k_2 < 0$ even if $\nu_1 > \nu_2$, 
thus the third field can be 
scattered opposite to the direction of propagation of the first two fields. 
} 
\end{figure}

Let us consider the four-wave mixing in a 3-level atomic medium. 
The pump and Stokes fields 
${\cal E}_1$ and ${\cal E}_2$ 
(whose Rabi frequencies are defined as $\W_1=\wp_1{\cal E}_1/\hbar$ and 
$\W_2=\wp_2{\cal E}_2/\hbar$, where $\wp_1$ and $\wp_2$ are the 
dipole moments of the corresponding 
transitions) 
with wave vectors $k_1$ and $k_2$ and angular frequencies $\nu_1$ and $\nu_2$ 
induce a coherence grating in the medium (see Fig.~\ref{4wm}) given 
by \cite{book} 
\be 
\rho_{cb} \sim -\W_1\W_2^*
\label{rcb} 
\ee 
Let us stress that the $\rho_{cb}$ coherence grating 
has an $\exp[i(k_1 - k_2)z]$ spatial dependence.  
In an ultra-dispersive medium (see Fig.~\ref{kw}) 
where fields propagate with a slow group velocity, 
the two co-propagating fields have wavevectors given by
\be
k_1 \simeq  k_1(\w_{ab}) + {\partial k_1\over\partial \nu_1}(\nu_1 - \w_{ab}) 
= \w_{ab}/c + (\nu_1 - \w_{ab})/V_g,
\label{k1k2}
\ee 
where $V_g$ is the group velocity of the first wave, 
$\w_{ab}$ is the frequency of transition between levels $a$ and $b$, 
and $k_2 = \nu_2/c$. Thus 
these two fields create a coherence grating in the medium 
with spatial phase determined by $k_1 - k_2 = \w_{cb}/c + (\nu_1 - 
\w_{ab})/V_g$ which depends strongly on the detuning $\delta = \nu_1
-\w_{ab}$.  By  properly choosing the detuning, $\delta$, 
one can make $k_1 - k_2$ negative. 
 
After the coherence $\rho_{bc}$ is induced in the medium, 
a probe field ${\cal E}_3$, with Rabi frequency $\W_3=\wp_3{\cal E}_3/\hbar$
and wave vector $k_3$, scatters off that coherence to produce 
the signal field, $\W_4$.  The signal field depends on the 
coherence and the input fields as 
\be 
{\partial \over\partial z}\W_4 \sim \rho_{cb}\W_3 \sim \W_1\W_2^*\W_3 
\sim e^{i(k_1 - k_2 + k_3-k_4)z} 
\ee 
That is, the propagation direction of $\W_4$ depends on 
the spatial phase of the $\rho_{bc}$ coherence 
through the phase-matching condition 
$k_4 = k_1 - k_2 + k_3$ \cite{boyd} while its frequency is determined by 
$\nu_4 = \nu_1 - \nu_2 + \nu_3$. 
 
We here show that for dispersive media one can obtain a strong signal in the 
backward direction even when all three input fields propagate forward. This is 
contrary to the usual non-dispersive media results, 
where the phase-matching in the backward direction 
cannot be achieved for ${\cal E}_1$, ${\cal E}_2$, and 
${\cal E}_3$ counter-propagating with respect to ${\cal E}_4$~\cite{boyd}. 
 
To demonstrate this result, we write 
the interaction Hamiltonian of the system as 
\beq 
V_I = -\hbar[\W_{2}e^{-i\w_{ac}t}|a\rangle\langle c| 
+\W_{1}e^{-i\w_{ab}t}|a\rangle\langle b| + h.c.]\\ 
-\hbar[\W_{3}e^{-i\w_{db}t}|d\rangle\langle b| 
+\W_{4}e^{-i\w_{dc}t}|d\rangle\langle c| + h.c.] 
\eeq 
where $\Omega_{4}=\wp_{4}{\cal E}_{4}/\hbar$ is the 
Rabi frequency of the signal field 
and $\w_{ab}$, $\w_{ac}$, $\w_{db}$, $\w_{dc}$ 
are the frequency differences between the corresponding 
atomic or molecular energy levels. 
The time-dependent density matrix 
equations are given by 
\be 
\frac{\partial{\rho}}{\partial{\tau}} 
=-\frac{i}{\hbar}[V_I, \rho]-\frac{1}{2}(\Gamma\rho+\rho\Gamma), 
\ee 
where $\G$ is the relaxation matrix. 
A self-consistent system also includes the field propagation equations 
\be 
\frac{\partial\W_1}{\partial{z}}=-i\eta_1\rho_{ab}, 
\;\;\; 
\frac{\partial\W_2}{\partial{z}}=-i\eta_2\rho_{ac}, 
\label{Maxwell1} 
\ee 
\be 
\frac{\partial\W_3}{\partial{z}}=-i\eta_3\rho_{db}, 
\;\;\; 
\frac{\partial\W_4}{\partial{z}}=+i\eta_4\rho_{dc}, 
\label{Maxwell2} 
\ee 
where 
\be 
\eta_j=\nu_j N \wp_{j}/(2\epsilon_0 c) 
\ee 
are the coupling constants $(j=1,2,3,4)$, 
$N$ is the particle density of the medium, 
$\epsilon_0$  the permitivity in vacuum. 
 
The equations of motion for the density matrix elements of the polarization
$\rho_{ab}$ and the coherence $\rho_{cb}$ are given by 
\be 
\dot{\rho}_{ab} = -\G_{ab}\rho_{ab} + i\W_1(\rho_{aa}-\rho_{bb}) -
i\rho_{cb}\W_2^*, 
\label{eq7} 
\ee 
\be 
\dot{\rho}_{cb} = -\G_{cb}\rho_{cb} + i\rho_{ca}\W_1 - i\rho_{ab}\W_2. 
\label{eq8} 
\ee 
where $\G_{ab} = \g_{ab} + i(\w_{ab}-\nu_1)$; 
$\G_{ca} = \g_{ca} - i(\w_{ac}-\nu_2)$; 
$\G_{cb} = \g_{cb} + i(\w_{cb}-\nu_1 + \nu_2)$; 
$\w_{cb}$ is the frequency of $c-b$ transition, 
and $\g_{\alpha\beta}$ are the relaxation rates 
at the corresponding atomic transitions. 
In the steady state regime, and assuming that  $|\W_2| >> |\W_1|$,
almost all of the population remains in the ground level 
$|b\ket$, $\rho_{bb}\simeq 1$. 
Let us consider the fields as plane waves: 
$\W_1(z,t) = \tilde\W_1(z,t) \exp(ik_1z)$, 
$\W_2(z,t) = \tilde\W_2(z,t) \exp(ik_2z)$, 
where $\tilde\W_1(z,t)$ and $\tilde\W_2(z,t)$ are the slowly varying 
in envelopes of the fields $\W_1$ and $\W_2$ in space, while 
$k_1=\nu_1[1 + \chi_{ab}(\nu_1)]/c$ and 
$k_2 = \nu_2[1+\chi_{ac}(\nu_2)]/c$. The succeptebilities are 
$\chi_{ab} = {\eta_1\rho_{ab}/\W_1}={(\nu_1 - \w_{ab})/ 2\pi V_g}$ and 
$\chi_{ac} = {\eta_2\rho_{ac}/\W_2}\simeq 0$. 
By solving 
the self-consistent system of Maxwell's equations
(\ref{Maxwell1},\ref{Maxwell2}) and the density matrix equations
(\ref{eq7},\ref{eq8}), we obtain Eq.~(\ref{k1k2}) for the wavevectors,  
where $V_g \simeq c/\eta_1|\W_2|^2$ is the group velocity of 
the optical field $\W_1$. 
Thus, the spatial dependence of $\rho_{cb}$ is determined by 
\be 
\Delta k = k_1 - k_2 = {\nu_1 - \nu_2\over c} + {\nu_1-\w_{ab}\over V_g}. 
\ee 
The signal field $\W_4$ is generated by the polarization $\rho_{db}$ of the 
transition it couples (Eq.~\ref{Maxwell2}). The equation of motion for this polarization element 
reads 
\be 
\dot{\rho}_{db} = -\G_{db}\rho_{db} + i\W_4(\rho_{dd}-\rho_{bb}) -
i\rho_{cb}\W_3^*,  
\ee 
where $\G_{db} = \g_{db} + i(\w_{db}-\nu_4)$, and 
$\nu_4$ is the frequency of generated field.
In the steady-state regime and for $|\W_4| << |\W_3|$, 
the field $\W_4$ at the output of the cell is given by 
\be 
\W_4 = \int_0^L dz e^{i(k_4 - \Delta k - k_3)z}{\eta_4\rho_{cb}\tilde\W_3^* 
\over\G_{db}} 
\label{int} 
\ee 
where $L$ is the length of the cell. 
Note here that Eq.(\ref{int}) is valid if the field $|\W_3|$ does not change
coherence $\rho_{cb}$ via power broadening which is true if 
$|\W_3|^2 \ll |\W_1|^2+|\W_2|^2$.

Hence, after  integrating 
Eq.(\ref{int}), we obtain for the scattered field $\W_4$ 
\be 
\W_4 = \ds\left[{\sin(k_4 - k_3 - \Delta k)L\over 
\ds(k_4 - k_3 - \Delta k)L}\right] 
{\eta_4 L\W_3\over\G_{db}} 
\label{W4} 
\ee 
where the expression in the brackets describes the phase-matching and 
determines the direction in which the signal field is generated. 

The most interesting effect following from Eq.(\ref{W4}) 
is the coherent backscattering. 
Indeed, even when all three input fields propagate forward, one may 
observe a backscattered signal field by satisfying the condition 
\be 
k_3 + \Delta k = k_3 + {\nu_1 - \w_{ab}\over V_g} < 0, 
\ee 
and for appropriate detuning, $\nu_1 - \w_{ab} < 0$, this inequality can be 
meet. That is, in order to obtain phase-matching in the backward direction, we
have to satisfy 
\be 
k_3 + \Delta k < -k_4, 
\ee 
which can be rewritten as 
\be 
-{\nu_1 - \w_{ab}\over V_g} > k_3 + k_4. 
\ee 
Hence, in order to demonstrate the effect, the detuning $\delta$ should meet 
the condition
\be 
\delta = - (k_3 + k_4) V_g \simeq - 2 k_4 V_g. 
\label{nu} 
\ee 
It is useful to rewrite the condition 
in terms of succeptibility for the probe field, indeed,
\be
k_1 = {\nu_1\over c} n_1 \simeq {\nu_1\over c} 
(1 + c{\nu_1-\w_{ab}\over \nu_1V_g}) =
{\nu_1\over c} (1 - c{2 k_4\over \nu_1}),
\ee
then, 
\be
\chi_{ab} = 2(n_1 - 1) =- 4{\lambda_{ab}\over\lambda_{db}},
\label{lambda}
\ee 
for gases $\chi_{ab} \ll 1$, so $\lambda_{ab}\ll\lambda_{db}$, i.e.
the effect can be implemented for scattering of IR fields. 
Then, for the Doppler broadened EIT media as shown in
\cite{matsko04oe,javan}, we can write 
\be
\chi_{ab}(\delta) \simeq {3\lambda_{ab}^3 N\over 8\pi^2}
\left({\g_r\delta\over|\W_2|^2}
+ i {\g_r\Delta_D\delta^2\over|\W_2|^4}
\right),
\ee
where $\Delta_D$ is the Doppler width;
$\g_r$ is the radiative decay rate. 
Thus, for detuning smaller than the EIT width
$|\delta| \leq \ds{|\W_2|^2/\sqrt{\g_r\Delta_D}}$, 
absorption can be neglected, and 
\be
{3\lambda^2_{ab} N\g_r\delta\over 16\pi|\W|^2} =-2 k_4, 
\ee
then, the atomic or molecular density is given by 
\be
N={32\pi k_4\over3\lambda_{ab}^2} 
\ds{|\W_2|^2\over\g_r|\delta|} \simeq 
{32\pi k_4\over3\lambda_{ab}^2} 
\sqrt{\Delta_D\over\g_r}.
\label{N}
\ee

\begin{figure*}[tb] 
\center{ 
\includegraphics[width=10cm]{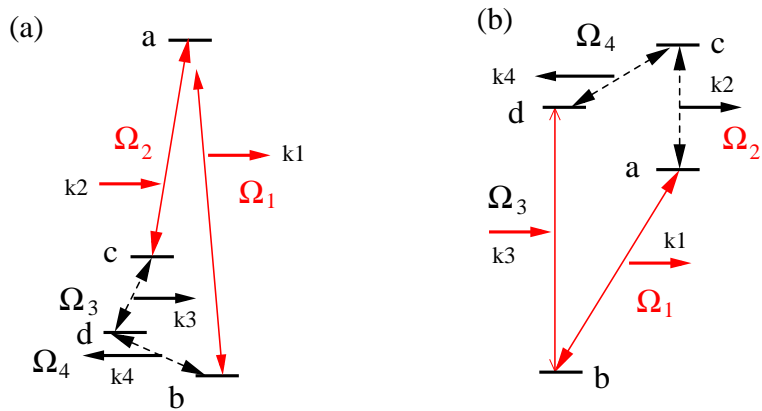} 
} 
\caption{\label{implementation} 
Implementation: molecular systems, 
(a) vibrational levels; 
co-propagating fields 1 and 2 
induce coherent between vibrational levels. 
The field 3 propagating in the same direction 
will be scattered in the opposite direction. 
(b) Atomic Rb scheme 
for implementation of coherent back scattering. 
} 
\end{figure*}

There are several schemes to demonstrate the effect.   
For example, the double-Lambda scheme can be implemented in molecular
rotational levels (see Fig.~\ref{4wm}a). Moreover, 
the effect can be implemented in the
ladder-$\Lambda$ using molecular vibrational levels (see
Fig.~\ref{implementation}a). The phase-matching condition should be 
slightly modified for the scheme as $k_4 = k_1 - k_2 - k_3$.    
Also, the phenomenon can be demonstrated in a V-$\Lambda$ scheme that 
can be realized in atomic levels
(see Fig.~\ref{implementation}b, for Rb atoms, $b=5S_{1/2}$, 
$c=7D_{3/2,5/2}$, $a=5P_{1/2,3/2}$, $d=8P_{1/2,3/2}$), and
phase-matching condition has a form $k_4 = k_1 + k_2 - k_3$. Let us note that
the requirement for detuning in all cases is $\delta/V_g = -2 k_4$ and 
the use of Eq.~(\ref{N}) to estimate molecular or atomic density is still valid. 

Examples of systems to this effect could be seen we mention
molecules $NO$ 
(a resonant transition at 236~nm, $A^2\Sigma^+-X^2\Pi$),
$NO_2$ (a resonant transition at wavelength 337~nm), 
and atomic Rb vapor (EIT and CPT have been recently
demonstrated for molecules, see \cite{eit-molecules}).  
The required molecular density of
$NO$ and $NO_2$ molecules is  $N\simeq 1.2\cdot
10^{13} \; \mbox{cm}^{-3}$ if one can use transition between rotational levels
$\simeq 10$ cm$^{-1}$. 
Using vibrational IR transitions for $NO$ (vibration frequency 
of 1900 cm$^{-1}$) at 5.26 $\mu$m
and for $NO_2$ (vibrational frequency of 750 cm$^{-1}$) 
13.3 $\mu$m, the densities are
$N=8\cdot 10^{15} \; \mbox{cm}^{-3}$ and 
$N=1.4\cdot 10^{15}$  $\mbox{cm}^{-3}$, correspondingly. 
For atomic Rb vapor, wavelengths are 
$\lambda_{1}=780$ nm, $\lambda_{2}=565$ nm,
$\lambda_{3}=335$ nm, $\lambda_{4}=23.4$ $\mu$m, 
and the atomic density is $N=1.4\cdot 10^{13} \; \mbox{cm}^{-3}$. 

The intensity needed for EIT is determined by condition
$|\W|^2\gg \g_{bc}\Delta_D$ which 
corresponds to laser intensity of the order of 1 mW/cm$^2$ for atoms and 
of the order of 10 W/cm$^2$ for molecules because the dipole moment is two
orders of magnitude smaller for molecules. 
These conditions are realistic and well-suited for an experimental
implementation.



Several applications of the effect can be envisioned, like in 
nonlinear CARS microscopy~\cite{xs-xie}, while the controlling of coherent
backscattering could provide a new tool for  
creating an image. A variation in the molecular density would modify 
the intensity of the signal in both the forward and the backward direction. 
Additionally, Eq.(\ref{W4}) also allows one to control the direction of the 
generated signal field and thus provide an all-optical control when scanning
an optical field over an object.  

In conclusion, 
we theoretically predict strong coherent scattering in 
the backward direction while using only forward propagating fields. 
This is achieved by exciting atomic or molecular coherence 
by properly detuned fields, in such a way that the resulting coherence has 
a spatial phase corresponding to a backward, counter-propagating wave. 
Applications of the technique to coherent scattering and 
remote sensing are discussed. The method holds promise for 
observation 
induced scattering in a backward direction with application to CARS 
microscopy. 
 
We thank J. Giordmaine, K. Hakuta, 
N. Kro\'o, G. Kurizki, K.K.~Lehmann, R. Miles, 
H. Walther, and X.S. Xie 
for useful discussions and gratefully acknowledge the support from 
the Defense Advanced Research Projects, the Office of Naval Research 
under Award No. N00014-03-1-0385, 
the Robert A.\ Welch Foundation (Grant \#A1261). 


\end{document}